# THE 1919 ECLIPSE RESULTS THAT VERIFIED GENERAL RELATIVITY AND THEIR LATER DETRACTORS: A STORY RE-TOLD

by

GERARD GILMORE FRS and GUDRUN TAUSCH-PEBODY*

*Institute of Astronomy, Madingley Road, Cambridge CB3 0HA, UK*

Einstein became world famous on 7 November 1919, following press publication of a meeting held in London on 6 November 1919 where the results were announced of two British expeditions led by Eddington, Dyson and Davidson to measure how much background starlight is bent as it passes the Sun. Three data sets were obtained: two showed the measured deflection matched the theoretical prediction of Einstein's 1915 Theory of General Relativity, and became the official result; the third was discarded as defective.

At the time, the experimental result was accepted by the expert astronomical community. However, in 1980 a study by philosophers of science Earman and Glymour claimed that the data selection in the 1919 analysis was flawed and that the discarded data set was fully valid and was not consistent with the Einstein prediction, and that, therefore, the overall result did not verify General Relativity. This claim, and the resulting accusation of Eddington's bias, was repeated with exaggeration in later literature and has become ubiquitous.

The 1919 and 1980 analyses of the same data provide two discordant conclusions. We reanalyse the 1919 data, and identify the error that undermines the conclusions of Earman and Glymour.

## INTRODUCTION

6 November 2019 was the centenary of the presentation to a joint meeting of the Royal Society and the Royal Astronomical Society of the scientific results from the May 1919 solar eclipse expeditions.[1] These determined that the reality and amplitude of gravitational light bending by the Sun were consistent with Einstein's 1915 predictions, and significantly inconsistent with the Newtonian flat-space-time prediction.[2] This evidence, complementing the earlier (1915) evidence that the theory explained an anomaly in the orbit of the planet Mercury, established General Relativity as a valid theory of space-time and made Einstein famous.[3]

*gil@ast.cam.ac.uk; gpebody@ast.cam.ac.uk

---

1 Sir Joseph Thomson, 'Joint eclipse meeting of the Royal Society and the Royal Astronomical Society', *Observatory* **42**, 389–398 (1919).

2 Sir F. W. Dyson, A. S. Eddington, C. Davidson, 'A determination of the deflection of light by the Sun's gravitational field, from observations made at the total eclipse of May 29, 1919', *Phil. Trans. R. Soc. Lond. A* **220**, 291–333 (1920).

The 1915 publication of Einstein's Theory of General Relativity included three definitive and necessary observational tests for which the General Relativistic prediction differed from flat space-time 'Newtonian' values. In the order presented by Einstein, the first is the rate of precession of the orbit of the planet Mercury around the Sun (an additional 43 arcsec per century, observation at the time required 45±5 arcsec per century). The second is the predicted amplitude of bending by a light-ray from a background star passing the Sun (1.75 arcsec at the Sun's limb). The third is the gravitational redshift of light for a particle moving through a gravitational potential; this last was determined unambiguously only in the 1960s.[4]

Einstein published the Mercury orbit calculation, and its match to observation on 18 November 1915, together with his prediction for the gravitational light bending. He then published his completed theory, 'The field equations of gravitation', on 25 November 1915. On 13 January 1916 Karl Schwarzschild published his exact (metric) solution of the field equations, confirming that Einstein's approximate calculation of the observational implications was accurate and the unique solution.[5] It was the determination that light bending matched the General Relativity prediction, derived from observations during the solar eclipse on 29 May 1919, made at Sobral, Brazil, and Principe, West Africa, and reported on 6 November 1919, which established wide support for the General Theory and made Einstein a global household name. The eclipse test involved three data sets, one of which was discarded as unreliable during the analysis, but which in isolation could be interpreted to produce a result discordant with Einstein's prediction. The other two established what at the time was accepted as a clear case preferring the 1.75 arcsec light-bending prediction over the smaller Newtonian value.

The 1919 published eclipse results were revisited in 1980, with a new interpretation of the evidence and claims about Eddington's bias, which were then further widely publicized and discussed in the context of a philosophical debate about science.

In a 1980 re-analysis of the published experimental data, philosophers of science John Earman and Clark Glymour presented their claim of a series of weaknesses in the 1919 analysis sufficient to undermine the original conclusions.[6] Their claim is twofold (if not always clearly so). They suggest that the discarded data set was of equal statistical validity to those retained and thus the resulting measurement published was not supportable by the data, and they also claim that the set (trichotomy) of three hypotheses proposed to be distinguished was inappropriate. They conclude that the result presented on 6 November 1919 in favour of General Relativity was determined not by evidence, but bias: '[Eddington's] bias showed in his treatment of the evidence: he repeatedly posed a false trichotomy for the deflection results, claimed the superiority of the qualitatively inferior Principe data, and suppressed reference to the negative Sobral results.' And further: 'This curious sequence of reasons might be cause enough for despair on the part of those who see in science a model of objectivity and rationality.'

---

3  Albert Einstein, 'Erklärung der Perihelbewegung des Merkur aus der allgemeinen Relativitätstheorie [Explanation of the perihelion motion of Mercury from the General Theory of Relativity]', *Preuss. Akad. Wiss. Sitzungsber.* **1915**, 831–839 (1915); Albert Einstein, 'Feldgleichungen der Gravitation [The field equations of gravitation]', *Preuss. Akad. Wiss. Sitzungsber.* **1915**, 844–847 (1915); Albert Einstein, 'Grundlage der allgemeinen Relativitätstheorie [The foundation of the General Theory of Relativity]', *Ann. Phys.* (Ser. 4) **49**, 769–822 (1916).

4  Clifford M. Will, 'The confrontation between General Relativity and experiment', *Living Rev. Relativity* **17**, 4 (2014) (http://dx.doi.org/10.12942/lrr-2014-4).

5  'Die Eindeutigkeit der Lösung hat sich durch die vorstehende Rechnung von selbst ergeben. Daβ es schwer wäre, aus dem Annäherungsverfahren nach Hrn. EINSTEINS Art die Eindeutigkeit zu erkennen, sieht man an folgendem: (…)': Karl Schwarzschild, 'Über das Gravitationsfeld eines Massenpunktes nach der Einsteinschen Theorie [On the gravitational field of a point mass according to the Einsteinian theory]', *Preuss. Akad. Wiss. Sitzungsber.* **1916**, 189–196 (1916).

6  John Earman and Clark Glymour, 'Relativity and eclipses: the British Eclipse expeditions of 1919 and their predecessors', *Hist. Stud. Phys. Sci.* **11**, 49–85 (1980). Their context from which we quote a summary is significant: 'This curious sequence of reasons might be cause enough for despair on the part of those who see in science a model of objectivity and rationality. That mood should be lightened by the reflection that the theory in which Eddington placed his faith because he thought it beautiful and profound—and, possibly, because he thought that it would be best for the world if it were true—this theory, so far as we know, still holds the truth about space, time and gravity' (at p. 85).

This claim was strongly supported in the popular 1983 book *The golem* (and elsewhere), where the authors' summary of the Earman and Glymour claim is now: 'Thus Eddington and the Astronomer Royal did their own throwing out and ignoring of discrepancies.'[7] In later literature, assertions that the eclipse results (often ascribed only to Eddington) were affected by bias, or were conscious cheating, multiply and are now ubiquitous, presented in a language increasingly more suggestive and less evidence-based.[8] In *Fabulous science*, historian John Waller concludes that Eddington 'serves to debase the whole scientific enterprise'.[9] Nickerson's review of confirmation bias includes 'Collins and Pinch's account of the reporting of the results of the 1919 expedition and of the subsequent widespread adoption of relativity as the new standard paradigm of physics represents scientific advance as somewhat less inexorably determined by the cold objective assessment of theory in the light of observational fact than it is sometimes assumed to be'.[10]

The measured gravitational deflection was supported by the results from the next available eclipse (Australia, 21 September 1922).[11] Evidence to support the validity of General Relativity as an accurate description of the Universe strengthened rapidly during and after the 1960s. What, then, motivated the doubts raised in the 1980s?

There are several open questions, which we consider in this article. What statistical significance, in modern terms, follows from re-analysis of the three sets of original eclipse data, and is the statistical re-analysis by Earman and Glymour actually valid? In this study we do both analyses, consider also Earman and Glymours's other criticisms and discover that, while the original 1919 analysis was statistically and methodologically robust, the 1980 re-analysis was not. We put this in a context contrasting scientific data analysis and philosophical debate. Where data analysis underpins an argument, it must be methodologically rigorous or there will indeed be cause for despair when objectivity and rationality are assumed as a standard.

---

7 Harry Collins and Trevor Pinch, *The golem: what everyone should know about science* (Cambridge University Press, Cambridge, 1993; 2nd edn 1998). There are many other supporting articles. For example, Alistair Sponsel, 'Constructing a "revolution in science": the campaign to promote a favourable reception for the 1919 solar eclipse experiments', *Br. J. Hist. Sci.* **35**, 439–467 (2002), has 'As John Earman and Clark Glymour persuasively demonstrated in their classic 1980 paper, the results themselves of the eclipse experiments were by no means an unequivocal confirmation of Einstein's theory.' David Deutsch, *The fabric of reality* (Penguin, London, 1997) has 'The observation of this deflection by Arthur Eddington in 1919 is often deemed to mark the moment at which the Newtonian world-view ceased to be rationally tenable. (Ironically, modern reappraisals of the accuracy of Eddington's experiment suggest this may have been premature).' As examples of a very large literature, see also Robert G. Hudson, 'Novelty and the 1919 eclipse experiments', *Stud. Hist. Phil. Mod. Phys.* **34**, 107–129 (2003), and reaction to that by Deborah G. Mayo, 'Novel work on problems of novelty?', *Stud. Hist. Phil. Mod. Phys.* **34**, 131–134 (2003).
8 Wikipedia, 'The Eddington experiment', https://en.wikipedia.org/wiki/Eddington_experiment (accessed 13 January 2020).
9 John Waller, *Fabulous science: fact and fiction in the history of scientific discovery* (Oxford University Press, 2004).
10 Raymond S. Nickerson, 'Confirmation bias: a ubiquitous phenomenon in many guises', *Rev. Gen. Psychol.* **2**, 175–220 (1998).
11 W. W. Campbell and R. Trümpler, 'Observations on the deflection of light in passing through the Sun's gravitational field, made during the total solar eclipse of September 21, 1922', *Publ. Astr. Soc. Pacific* **35**, 158 (1923); W. W. Campbell and R. J. Trümpler, 'Observations made with a pair of five-foot cameras on the light-deflections in the Sun's gravitational field at the total solar eclipse of September 21, 1922', *Lick Obs. Bull.* **13**, 130 (1928). It is interesting that it was only after receiving the results of the 1922 eclipse that Dyson finally accepted that there was no unrecognized systematic error vitiating the 1919 results. Another British astronomer, Harold Spencer Jones, commented: 'It is no longer open to [opponents of Einstein's Theory] to say that the prediction of the amount of the deflection has not been fully confirmed.' This story is told in detail in Jeffery Crelinsten, *Einstein's jury: the race to test relativity* (Princeton University Press, 2006) at p. 213.

There is a substantial published literature reassessing the roles and responsibilities of those involved in the eclipse. Of specific direct relevance is a series of investigations by Daniel Kennefick.[12] Kennefick demonstrates robustly that the two expeditions' plates were reduced and analysed separately. The Sobral plate analyses, including the decision to exclude the Sobral astrograph results, was made by the Royal Greenwich Observatory team under Dyson. Eddington worked independently. Kennefick, and other studies, make the case convincingly that attributing the data set selection to Eddington is not historically accurate. Indeed, Kennefick concludes that there are good grounds for believing that Dyson made a scientifically correct decision in choosing to reject the astrographic data, based on Dyson's well-advertised reserve about unquantifiable systematic errors.

Many other studies have addressed the same or related questions, often concluding that the 1919 eclipse analysis teams did show objective integrity and base their conclusions on accepted standards of scientific evidence. Examples are Ben Almassi, Matthew Stanley, Jeffrey Crelinsten and Jean Eisenstaedt, none of whom adopts the Earman and Glymour claims uncritically, and all of whom make strong cases for the integrity of the eclipse teams.[13] Other authors continue to endorse or simply to adopt the Earman and Glymour conclusions (eg Schindler, Miller, Bolinska and Martin;[14] none of these studies, however, explains how and why Earman and Glymour reached their clearly influential conclusion and examines the correctness of the scientific evidence they present. We do that in this study, and return to the sources of the argument.

---

12 Daniel Kennefick, 'Testing relativity from the 1919 eclipse: a question of bias', *Phys. Today* **62**, 37–42 (2009); Daniel Kennefick, 'Not only because of theory: Dyson, Eddington and the competing myths of the 1919 eclipse expedition' in *Einstein and the changing world views of physics* (eds Christoph Lehner, Jürgen Renn and Matthias Schemmel), pp. 201–232 (Birkhaüser, Basel, 2012); and Daniel Kennefick, *No shadow of a doubt: the 1919 eclipse that confirmed Einstein's Theory of Relativity* (Princeton University Press, 2019). A positive comment on Kennefick's discussion is Philip Ball, 'Arthur Eddington was innocent!', *Nat. Online* (2007) (doi:10.1038/news070903-20), where Ball concludes: 'The motto of the Royal Society—*Nullius in verba*, loosely translated as "take no one's word"—is often praised as an expression of science's guiding principle of empiricism. But it should also be applied to tellings and retellings of history—we shouldn't embrace cynicism about how scientists do their work just because it's become cool to knock historical figures off their pedestals.' See also Matthew Stanley, *Einstein's war* (Penguin Books, London, 2019).

13 Contemporary acceptance of the 1919 results has been made clear many times. For example, Ben Almassi quotes all of Earman and Glymour, Collins and Pinch, and Waller at face value. However, he continues: 'Such seemingly suspicious considerations might lend credence to Waller's allegations of intellectual dishonesty. But if we look carefully at the reasons Eddington and his colleagues gave in defence of their data analysis, where and to whom these reasons were presented, and the actual critical response from contemporary observers, I suggest, the case for this allegation is seriously undermined. … The fact that British physicists and astronomers were capable of critically engaging those parts of Eddington's work which have since been labelled contentious, and the fact that those who publicly disputed Eddington's claim to have confirmed relativity did so on different grounds, suggest that while Eddington's expertise played a crucial role in this case, the widespread acceptance of his expert testimony was not entirely without corroboration.' Ben Almassi, 'Trust in expert testimony: Eddington's 1919 eclipse expedition and the British response to General Relativity', *Stud. Hist. Phil. Mod. Sci.* **40**, 57–67 (2009); Matthew Stanley, 'An expedition to heal the wounds of war: the 1919 eclipse and Eddington as Quaker adventurer', *ISIS* **94**, 57–89 (2003) makes the same point. See also Matthew Stanley, *Practical mystic: religion, science and A. S. Eddington* (University of Chicago Press, 2007); Jean Eisenstaedt, *The curious history of relativity* (Princeton University Press, 2006); Jeffrey Crelinsten, 'William Wallace Campbell and the "Einstein problem": an observational astronomer confronts the Theory of Relativity', *Hist. Stud. Phys. Sci.* **14**, 1–91(1983).

14 'Kennefick's defense of the British eclipse physicists fails: his arguments are simply ineffective against Earman and Glymour's critique': Samuel Schindler, 'Theory-laden experimentation', *Stud. Hist. Phil. Sci.* **44**, 89–101 (2013). Many authors continue to quote Earman and Glymour, Collins and Pinch, and Waller at face value, as established studies: Boaz Miller, 'Why knowledge is the property of a community and possibly none of its members', *Phil. Q.* **65**, 417–441 (2015); Agnes Bolinska and Joseph Martin, 'Negotiating history: contingency, canonicity and case studies', *Stud. Hist. Phil. Sci. A* **80**, 37–46 (2020) (doi:10.1016/j.shpsa.2019.05.003).

THE 1919 ECLIPSE EXPEDITION AND ITS RESULTS

The importance of astronomical verification of the predictions of Einstein's General Theory of Relativity were emphasized by Einstein from the time he developed the Principle of Equivalence (1907) and realized that light bending was a consequence—with the prediction being published in 1911. At that time Einstein started a collaboration with Erwin Finlay Freundlich to encourage astronomers to search for the light-bending effect. In his presentations of the complete version of General Relativity in late 1915 and the overview paper in 1916, Einstein presented his calculations of the three experimental tests. These were the rate of precession of the perihelion of the planet Mercury, light bending close to the Sun and the gravitational redshift. He showed the Mercury orbit correction was in excellent agreement with observation, solving a long-standing problem. He repeatedly emphasized that there was no freedom in his theory to adjust the predictions (no free parameters), so experimental test of the other two predictions would support or refute the theory.

The predicted light-bending amplitude in General Relativity at the limb of the Sun is 1.75 arcsec, with deflection radial from the centre of the Sun and decreasing linearly with distance. The predicted light bending in a theory in which light responds to gravity, that is the equivalence principle applies but space-time is not curved, is exactly one-half of this value: 0.87 arcsec. This model was referred to at the time as the Newtonian prediction. A third possibility was considered, of zero deflection. Thus, the null hypothesis was to distinguish between these three possibilities.[15]

In this context, several observatories with solar eclipse expertise re-examined historical photographic material and/or organized dedicated expeditions to eclipses in 1912, 1914 and 1918. None succeeded in obtaining published results. The next available eclipse, on 29 May 1919, was exceptional in that the eclipse totality would be a relatively long duration, while the background star field was the Hyades star cluster, providing an exceptional number and spatial distribution of bright stars. Astronomer Royal Sir Frank Dyson recognized the unprecedented opportunity this provided to test light bending, and supported organization of a dedicated expedition that 'should serve for an ample verification, or the contrary, of Einstein's theory'.[16]

---

15  The early history of the calculation of Newtonian light deflection is presented in Clifford M. Will, 'Henry Cavendish, Johann von Soldner and the deflection of light', *Am. J. Phys.* **56**, 413 (1988). We note that a devotee of the wave theory of light would consider zero deflection the Newtonian value. Coincidentally, the person who recognized Cavendish's light-bending work was the same Dyson involved in the 1919 eclipse. On Soldner, see also Stanley L. Jaki, 'Johann Georg von Soldner and the gravitational bending of light, with an English translation of his essay on it published in 1801', *Found. Phys.* **8**, 927–950 (1978), and Jean Eisenstaedt, 'The Newtonian theory of light propagation', in *Einstein and the changing world views of physics* (eds Christoph Lehner, Jürgen Renn and Matthias Schemmel), pp. 201–232 (Birkhaüser, Basel, 2012). An important history and correction of some popular mis-statements about light bending, and gravitational lensing, is provided by David Valls-Gabaud, 'The conceptual origins of gravitational lensing', in Albert Einstein Century International Conference, *AIP Conf. Proc.* **861**, 1163–1171 (2006).

16  Sir F. W. Dyson, 'On the opportunity afforded by the eclipse of 1919 May 29 of verifying Einstein's Theory of Gravitation', *Mon. Not. R. Astr. Soc.* **77**, 445–447 (1917). It is worth noting just how exceptional is the 29 May 1919 eclipse. There are several bright stars suitably distributed near the solar limb, while the eclipse is of relatively long duration. It is the combination of both these factors that allowed the test of both light bending and the 1/*r* deflection law, illustrated in the eclipse report Diagram 2. Later eclipses

Two expeditions, both British, were eventually sent; one to Sobral (Brazil) staffed by Charles Davidson and Andrew Crommelin, both astronomers working at the Royal Greenwich Observatory, and one to Principe, an island off West Africa, led by Arthur Eddington from Cambridge, supported by Edwin Cottingham, a Cambridge clockmaker. No Cambridge astronomy staff were available, both assistants at Eddington's observatory having been killed during the war. Three telescopes were taken, two to Sobral, an astrograph and a 4-inch, and one astrograph to Principe. All were mounted horizontally under sun-shelters, with the view of the sky fed in by a moving coelostat mirror mounted in front of each telescope. It was recognized prior to the eclipse expeditions that the coelostat mirrors, and how they would respond to temperature changes during the eclipse, were a risk to be managed. A full description of both expeditions, their equipment and the subsequent reduction and analysis of their results was published in the eclipse report.

The principle of the experiment is straightforward. The astronomer obtains images of the star field near the Sun during solar eclipse. These are compared with photographs of the same star field taken by the same telescope some months earlier or later, when the stars can be photographed at night-time. The small differences in relative position of each star image are analysed to determine changes in scale and orientation between the two photographs, and any systematic effect caused by gravitational light bending. For stability, the photographs use glass plates and are typically referred to as plates. The telescope need not be kept in location at the eclipse site for the several months—rather, a sanity check on possible changes is made by comparing night-time images on 'check plates' of a different star field taken at both locations. In practice, the Sobral team identified significant coelostat-induced image and field distortion (astigmatism) with their astrograph, and so committed to waiting in Brazil until July to provide comparison plates with consistent distortion. Comparison plates for the Principe astrograph were obtained earlier from Oxford, its home observatory, and check plates were taken at Principe and Oxford.

The sky was mostly clear during totality in Sobral, so the eclipse team obtained 16 useful plates with the astrographic telescope and seven useful plates with the 4-inch telescope. All plates taken with the 4-inch telescope had satisfactory image quality. Those with the astrograph were, however, 'diffused and apparently out of focus. Worse still this change was temporary', the unadjusted telescope being back in focus for the evening check and July comparison plates. The eclipse was affected by thick cloud at Principe, with eventually only two plates proving satisfactory for analysis. All relevant differential star measurements from all three data sets are published in the eclipse report, allowing astronomers of the time, and us, to repeat the calculations.[17]

---

had fainter and more distant stars, and/or less uniform spatial distributions, making measures more dependent on small deflections, and the whole experiment more technically challenging. The very asymmetric star distribution in the 1929 eclipse was one major factor in the arguments about its data analysis—see appendix B.

17   For example, Henry Norris Russell, 'Note on the Sobral eclipse photographs', *Mon. Not. R. Astr. Soc.* **81**, 154–164 (1920). In a set of three papers, Louis A. Bauer examined the 1919 eclipse results, including his own geophysical expedition, and quoted the Lick/Goldendale preliminary result from 1918, even though that was never published as a formal result. His particular focus is on the many possible atmospheric phenomena that were also considered in the UK discussions. Louis A. Bauer, 'Résumé of observations concerning the solar eclipse of May 29, 1919, and the Einstein effect', *Science* **51**, 301–311 (1920); Louis A. Bauer, Preliminary results of analysis of light deflections observed during solar eclipse of May 29, 1919, *Science* **51**, 581–585 (1920); Louis A. Bauer, Further results of analysis of light deflections observed during solar eclipse of May 29, 1919, *Science* **52**, 147 (1920). In spite of many statements to the contrary, the original photographic plates from the Sobral telescopes remain in storage as part of the UK National Archive, managed by the Bodleian Library. Most, but not all, plates were re-measured and re-analysed in 1979, as part of the Einstein centenary; G. M. Harvey, 'Gravitational deflection of light: a re-examination of the observations of the solar eclipse of 1919',

Table 1. Comparison of the agreements of each telescope with the three theoretical hypotheses. The $P(z)$ values are the probabilities of observing a result at least as extreme as the prediction due to the Einsteinian, Newtonian or null (no deflection) model (in either tail) of a normal distribution with mean and standard deviation of the appropriate telescope. So, for example, the results from the Principe astrograph are consistent with the Newtonian prediction with 10% probability, while the Sobral 4-inch results are consistent only with the Einstein deflection, having less than 1 in 1000 chance of agreeing with the Newtonian prediction. $z$ is the difference between the observed and predicted values in units of standard deviations.

| Telescope analysis | Deflection (arcsec) | Einstein | | Newton | | Null | |
|---|---|---|---|---|---|---|---|
| | | $z$ | $P(z)$ | $z$ | $P(z)$ | $z$ | $P(z)$ |
| Sobral 4-inch | $1.98 \pm 0.18$ | 1.28 | 0.201 | 6.17 | 0.000 | 11.0 | 0.000 |
| Principe astrograph | $1.61 \pm 0.45$ | $-0.31$ | 0.756 | 1.64 | 0.100 | 3.58 | 0.000 |

The report describes analysis of each of the three data sets, one per telescope. Considerable care is made to clarify the importance of uncertainties, with contributions from both random and systematic errors. We discuss the numerical work, and our re-analysis of the data, in appendix A. The important point for the current discussion is that considerable care was taken to explain the limits of validity of the results, and hence the significance of the conclusions.

The Sobral 4-inch plates gave the most precise and accurate result, with the derived result for light bending, corrected to the limb of the Sun, of $1.98 \pm 0.12$ arcsec. The error quoted is the probable error. Converting to the now common standard deviation gives $1.98 \pm 0.18$ arcsec. The Principe results were based on only two satisfactory plates, and are hence less precise. Eddington derived a final value of $1.61 \pm 0.30$ arcsec (probable error), or $1.61 \pm 0.45$ arcsec (standard deviation). The eclipse report emphasizes the large contribution to this final uncertainty from possible systematic errors. For the Sobral astrograph, no final derived limb deflection was presented. Rather, the formal result quoted was 0.93 arcsec, 'discordant by an amount much beyond the limits of its accidental error'. Two possible results were in fact presented for the Sobral astrograph analysis, 0.93 arcsec and 1.52 arcsec, neither with a quoted uncertainty and both depending on assumptions made as to the cause of the distorted eclipse images. The report discards consideration of these data (see appendix A for further technical discussion).

The accepted data provide a significance test (in modern terms) of the null hypothesis (see appendix A for more detail). Table 1 presents the significance test results. This makes clear that the Sobral 4-inch data are highly inconsistent with the Newtonian and null possibilities and consistent with Einstein, while the Principe data are consistent with Einstein and mildly consistent with Newtonian, but not with null. We combine these into a single experimental outcome in appendix A.

Accepting rejection of the Sobral astrograph data, the evidence available in the eclipse report does indeed support the conclusion that 'the results of the observations here

---

*Observatory* **99**, 195 (1979). The Principe plates have been lost. Harvey does not re-analyse the exact same set of stars and plates used by the 1919 report, so his conclusions in detail are not exactly comparable. His results do confirm that the star positions were reliably determined in the 1919 measures. We restrict attention here to the original 1919 measurements.

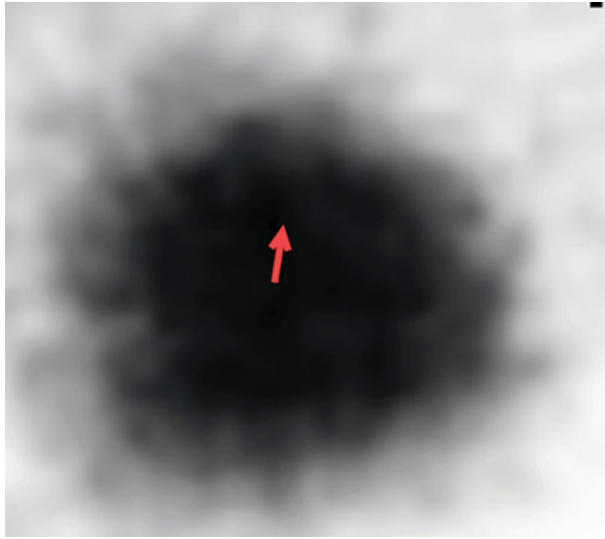

Figure 1. The image is an enlargement of 'Star 4' from a Sobral 4-inch plate. The red arrow indicates the amplitude of the centroid shift due to gravitational deflection at the position of this star, which is 0.75 arcsec. This image is from a PDS (Photometric Data Systems) machine scan of an eclipse plate made by John Pilkington at the Royal Greenwich Observatory in 1999, at the request of and kindly provided by Dr Robin Catchpole.

described appear to point quite definitely to the third alternative, and confirm Einstein's generalised relativity theory'.

It is important for this article to note how much those involved emphasized the importance of systematic errors affecting the Sobral astrograph data. The eclipse report says the individual measures were 'discordant by an amount much beyond the limits of accidental error'. In his meeting summary for *Nature*, Crommelin emphasizes: 'The probable error, as estimated by the individual discordances, is about 0.3 arcsec, but there is reason to suspect systematic error, owing to the very different character of the star images on the eclipse and check plates.'[18] In 1921 Dyson says about the measurements that 'these displacements, though small, are ten times as large as those met with in determinations of stellar parallax' (cf. figure 1).[19] He repeats the results, for the Sobral astrograph, as 0.90 arcsec if one assumes a scale change, whereas if one assumes just a distortion the result is 1.56 arcsec (*sic*, not 1.52 arcsec as in the report), 'a result of little weight', and with no final error presented.

It is also important to clarify here that there was no widely accepted disagreement at the time with the validity of the observational determination of the deflection. This point is significant to establish that Earman and Glymour were not building on evidence contemporary to the 1919 discussion. To avoid distracting from the main focus of this article, we briefly review that evidence, together with the 1980s context, in appendix B.

---

18  A. C. D. Crommelin, 'Results of the total solar eclipse of May 29 and the Relativity Theory', *Nature* **104**, 280–281 (1919).
19  Frank Dyson, 'Relativity and the eclipse observations of May, 1919', *Nature* **106**, 786–787 (1921). Bauer, *op. cit.* (note 17) is also relevant here.

## The Earman and Glymour (1980) re-analysis

In 1980 John Earman and Clark Glymour re-analysed the results of the 1919 eclipse. They reached strong conclusions: 'One may imagine that in order to turn the tide of opinion the eclipse results must have been unequivocal. They were not.' 'The British results, taken at face value, were conflicting, and could be held to confirm Einstein's theory only if many of the measurements were ignored. Even then, the value of the deflection obtained was *significantly* higher than the value Einstein predicted' (our emphasis—we return to this misuse of technical statistics terminology below).

We examine first the claim about conflicting results. In their article, Earman and Glymour re-consider the evidence, looking specifically at the uncertainties. They quote Eddington's Principe limb-deflection value from the eclipse report, correcting to standard deviation errors, as 1.61 ± 0.444 arcsec, noting that this scaling assumes normal (error) distributions, and noting the contribution from systematic error 'including contributions from the error in scale determination'. They hint at a possible limitation in Eddington's analysis, but do not do any calculations to support the doubt raised: 'It is possible that … Eddington reached a final estimate of the gravitational deflection slightly different from what a straightforward least-squares determination would have given.' We have carried out the least-squares determination (see appendix A) and confirm Eddington's calculation as valid. There is no cause for doubt.

Earman and Glymour then provide a value for the equivalent standard deviation for the Sobral astrographic dataset, being 0.48 arcsec, even though no such uncertainty was quoted in the report. They state: 'the dispersion in the measurements from the Principe astrographic is about the same as the dispersion from the Sobral astrographic. … In all, these sets of measurements seem of about equal weight, and it is hard to see decisive grounds for dismissing one set but not the other.' The lack of precision in this statement, highlighted by the absence of calculations, an unfortunate feature of Earman and Glymour's analysis, is clarified further below, while an immediate question here is the origin of the number quoted for the Sobral astrograph standard deviation. No uncertainty was quoted in the report. Indeed, as noted above, the report and all presentations of the eclipse data at the time repeatedly emphasize that there is no robust answer, that either 0.93 arcsec or 1.52 arcsec is a possible mean result, but that no formal uncertainty can be derived because the systematic error is so large. Additionally, the report makes explicit the evidence that the results from the Sobral astrograph are not normally distributed. Where did Earman and Glymour's quoted standard deviation come from? The explanation is apparent on their page 75, where they switch from use of the term 'standard deviation' to the term 'dispersion'. Indeed, the value they quote, 0.48 arcsec, is the internal dispersion of the 18 results of the 16 measured Sobral astrograph plates, provided one naively treats the dispersion as representing an underlying normal distribution. Earman and Glymour then consider this internal dispersion as equivalent to a standard deviation-like uncertainty on the mean, which would include a contribution from systematic errors. This is a conceptual and mathematical error. For comparison, the internal dispersion of the Principe astrographic data is 0.22 arcsec, and the total uncertainty on the mean is 0.45 arcsec, not at all comparable to the Sobral value. Earman and Glymour confuse internal dispersion with standard deviation of a mean, and neglect the contribution of systematic errors in the case most emphasized to be affected by them. The numbers they quote are not comparable, as we discuss further in appendix A (table A2). Their statement that 'the dispersion in the measurements from the Principe astrographic is about the same as the dispersion from the Sobral astrographic' is a conceptual and mathematical error that inevitably undermines the validity of all the conclusions they later derive from it.

Conceptual error similarly affects Earman and Glymour's claim that 'the very quality of the Sobral 4-inch results prevent them from constituting an unequivocal confirmation of Einstein's predicted deflection of 1.74 arcsec: the mean value is too high and the dispersion too small'. The 4-inch result, as noted in table 1, is only 1.3 standard deviations from the Einstein prediction, fully consistent with agreement. It is mathematically incorrect to conclude that the 4-inch result is 'significantly above even Einstein's value'.

Earman and Glymour then speculate that 'If one kept the data from all three instruments, the best estimate of the deflection would have to be somewhere between the Newtonian value and the Einstein value'. We have carried out the missing calculations (detail in appendix A, table A4), and show Earman and Glymour's guess is incorrect. All ways of combining the data, following standard statistical weighting techniques, provide a combined weighted result slightly, but not significantly, larger than Einstein's value.

In a paired argument against the robustness of the eclipse report's conclusion, Earman and Glymour claim the three-way theory choice ('trichotomy') adopted by Eddington was inadequate—Eddington 'repeatedly posed a false trichotomy for the deflection results'. This claim, while widely repeated, even in current literature, is surprising. The theories available at the time were: (1) no light deflection (scalar-theory gravitation); (2) light has weight, and space-time is Euclidean (Newtonian); or (3) light has weight, and space-time is curved (General Relativity). Einstein emphasized that there were no free parameters involved in his calculations, so there was no other possibility with strong theoretical justification at the time, and hence the experimental test was crucial.

The three-way test applied to the eclipse results is in fact a typical example of what is routinely called today in the scientific community a 'null hypothesis test'. This is a common and very widely applied methodology when testing a new but not high-precision experimental result against possible available theories. A recent example is the first image of a super-massive black hole. The experiment was not anticipated to be able to provide high-precision tests of the null hypothesis, in this case that a black hole is described by the Kerr metric and the no-hair theorem, but could provide some useful limits of validity. In the 1919 case, the differences between the three available hypotheses were large enough that a distinction between them was indeed correctly anticipated. The null hypothesis test is the test still widely applied in similar circumstances by scientists. While Earman and Glymour may consider that 'false', it was and remains the way science and scientists operate.

Questioning the choice of three hypotheses that the experiment was set to test, as Earman and Glymour do in this case, is not the same as questioning the scientific validity of results within the parameters defined by the choice of hypotheses. The first is a philosophical argument, the second a scientific one. The advantage of keeping this distinction clear becomes evident when we consider the references to Earman and Glymour's article in later literature, particularly in the light of another point or question sometimes conflated in discussion about the trichotomy; that is, the determination of how precisely the predictions of one theory (General Relativity) can be tested by experiment.

After the early successes of eclipse testing, later eclipses attempted to add more significant digits to the measurement accuracy, without success.[20] This was, however, no longer null

---

20  Harald von Klüber, 'The determination of Einstein's light-deflection in the gravitational field of the Sun', *Vistas Astron.* **3**, 47–77 (1960), provides a convenient uncritical compilation of all published eclipse results up to 1960, collecting reports from a variety of obscure or rare sources. Nonetheless, von Klüber does spend most of his article repeating the arguments of Freundlich, von Klüber and von Brunn that all results should be treated very sceptically since they do not implement the recommended Potsdam absolute hypothesis testing in the 1919 sense. It was attempting to answer the question 'how accurate is General Relativity?', which is different from the question 'can we distinguish between the null/ Newtonian/ General Relativity predictions?' Each of these questions can be subject to test by the same experiment, but only the answer to the first requires the precision measurement that is sometimes used by commentators to invalidate the answer to the second—see our appendix B for more comment.

Conceptual and mathematical errors, and speculation, rather than calculation and factual evidence, undermine the argument emerging from Earman and Glymour's analysis that neither the astronomers who presented the data on 6 November 1919 nor their audience were convinced by the robustness of the results verifying Einstein's theory. And it is this belief that leads then to their assertion—presented as fact on their page 79—that 'Eddington won the argument by the power of the reference work'. Little consideration is given to the historical evidence, presented on their page 78, that 'the account of the proceedings records no queries about the probable error of the Sobral 4-inch determination, or about the warrant for giving some weight to the Principe results but none to those of the Sobral astrographic'. This statement extends beyond the meeting report to the extensive discussions of the results in the literature in subsequent years. The audience to whom the results were presented on 6 November 1919 was an audience of experts, well versed in the conceptual distinctions Earman and Glymour fail to make, and proficient enough mathematicians to assess for themselves the robustness of what was put in front of them— if not immediately, then certainly by the time the eclipse report with full results was published (see also appendix B).[21]

Earman and Glymour make one other point, which is important to put into context not only their analysis but also the use of their arguments in later literature. They raise the (presentist) possibility that the method used by Einstein (and by Eddington) to calculate the light deflection (and the precession of the planet Mercury) from Einstein's field equations was 'from a modern point of view … problematic', thus raising the question whether the Einstein value had been robustly established prior to the experiment. Note that the authors merely flag this up as a possibility, but do not use it to reinforce their other arguments. It is only in the later book, *The golem*, which draws heavily and not always critically on Earman and Glymour's article, that this point is picked up and exaggerated to a very major criticism. The authors of *The golem* claim to have found here an example of an iterative prediction–observation mutual reinforcement: 'In interpreting the observations this way, Eddington seemed to confirm not only Einstein's prediction about the actual displacement, but also *his method of deriving the prediction from his theory*—something that no experiment can do' (their emphasis).[22]

The detailed history of the Einstein–Besso method, used by Einstein to calculate the three classical predictions of General Relativity in November 1915, has become widely studied since discovery of key documents in the Einstein papers in the 1980s; there are several

---

calibration method. A non-expert reader could only conclude that great reserve should be applied to all results. A significant factor in this is that no later eclipse enjoyed the exceptionally favourable background star field that facilitated the results of the 1919 eclipse. See appendix B and note 16.

21  It is of relevance here that Eddington presented the eclipse results on 22 October 1919 in a talk to a special meeting of the Cambridge 'Grad-squared V' society. The 27 members who discussed Eddington's talk, 'The weight of light', included the elite of Cambridge physicists, mathematicians and astronomers. The positive reception of his results here seems to have reassured Eddington that they were ready for public release. The relevant minute book is Cambridge University Library, University/SOC.XXIX.1.

22  Collins and Pinch, *op. cit.* (note 7), at p. 45.

recent, comprehensive and accessible studies.[23] That this calculation technique is complex and therefore was not easy to understand by the community of the time was emphasized by Nobel laureate and President of the Royal Society J. J. Thomson, who refers in his report of the announcement meeting to the 'great difficulty' of expressing the new theory mathematically: 'It would seem that no one can understand the new law of gravitation without a thorough knowledge of the theory of invariants and of the calculus of variations.' That is not to say, however, that Einstein's calculations had not been understood or verified. As early as January 1916 the exact Schwarzschild metric solution was published. In this paper, Schwarzschild explicitly showed that the Einstein calculations were both unique solutions and sufficiently accurate for the calculations. That same year de Sitter, referring to both the Schwarzschild and Droste solutions, provided full detail of the derivation of the deflection of light. De Sitter's review, the first English-language presentation of General Relativity, discusses all this in two articles.[24] In §12 of his first article, he notes that the solar limb deflection is 1.75 arcsec, which 'could probably very well be measured'. There is therefore no reason to suspect that the difficulty of Einstein's calculations had been in the way of establishing robustly the value to be tested.

Having established (incorrectly) that the results of the British 1919 expeditions were not robust and worthy of the recognition they received from the scientific community of the time and beyond, Earman and Glymour presented alternative explanations of the expeditions' historical success, introducing generalization and allegation, and speculating about motivation—all of which were misleading, especially when they were picked up in later literature. The comments Earman and Glymour made, and their interpretation of the facts they chose to refer to are based on the assumption that the scientists involved in the expeditions were determined to show that their results verified Einstein's Theory of General Relativity in spite of what the authors allege were flawed data. The claim that the scientists involved, Eddington in particular, were biased to the extent of falsifying experiment is, as far as we are aware, introduced here for the first time.

According to Earman and Glymour: 'Dyson spoke as the voice of both expeditions, and it is striking that he described the Principe expedition but did not mention its results. It appears that Dyson had decided to discount the results of both the Sobral and the Principe astrographics.'[25] This is a puzzling summary of the meeting report, where Dyson provides a brief motivation for the whole project, some specifics of the telescopes borrowed and an overview of the activity at Greenwich. Dyson did not 'describe the Principe expedition', nor is there evidence in the meeting report to deduce anything about his opinion of the Principe results other than that he agreed to co-author their publication. Nonetheless, Earman and Glymour reach firm unreserved conclusions: '[Eddington's] bias showed in his treatment of the evidence: he repeatedly posed a false trichotomy for the deflection results, claimed the superiority of the qualitatively inferior Principe data, and suppressed reference to the negative Sobral results.'[26]

---

23  Eisenstaedt, *op. cit.* (note 13); H. Gutfreund and J. Renn, *The formative years of relativity: the history and meaning of Einstein's Princeton lectures* (Princeton University Press, 2017); Jürgen Renn and Tilman Sauer, in *Einstein's Zurich notebook: introduction and source* (ed. M. Janssen, J. D. Norton, J. Renn, T. Sauer, J. Stachel and L. Divarci), p. 113; Jürgen Renn (ed.), *The genesis of General Relativity* (Springer, Dordrecht, 2007), vol. 1, especially §7.17, pp. 280ff. See also J. Earman and M. Janssen, *The attraction of gravitation: new studies in the history of General Relativity* (Birkhaüser, Boston, 1993), at p. 129. Note their footnote 1.

24  W. de Sitter, 'On Einstein's theory of gravitation and its astronomical consequences: first paper', *Mon. Not. R. Astr. Soc.* **76**, 699–728 (1916), at §12, pp. 717–719, and Erratum, *Mon. Not. R. Astr. Soc.* **77**, 481 (1917); W. de Sitter, 'On Einstein's theory of gravitation and its astronomical consequences: second paper', *Mon. Not. R. Astr. Soc.* **77**, 155–184 (1916).

25  Earman and Glymour, *op. cit.* (note 6), p. 75.

Earman and Glymour conclude: 'Now the eclipse expeditions confirmed the theory only if part of the observations were thrown out and the discrepancies in the remainder ignored; Dyson and Eddington, who presented the results to the scientific world, threw out a good part of the data and ignored the discrepancies. This curious sequence of reasons might be cause enough for despair on the part of those who see in science a model of objectivity and rationality.'[27]

Our re-analysis shows that these strong claims are based entirely on methodological error. Earman and Glymour failed to understand the difference between the dispersion of a set of measurements and an uncertainty, random plus systematic, on the value of the parameter being measured. They speculated but did not calculate, and their conclusions are not supported by evidence.

Their error was left unchallenged and the strong conclusions and accusations they derived from it were used not only to question the scientific method then applied, but also to undermine the scientific integrity and reputation of an eminent scientist.

## Influence and consequence

There is a vast literature about the 1919 expeditions, with more publications appearing recently for the centenary of the eclipse. We discuss the directly relevant literature in appendix B. Two very recent overview studies of particular note are by Daniel Kennefick and Matthew Stanley.[28] Both works include discussion of the accusations of bias made against the scientists involved in the 1919 eclipse expeditions; both mention the strength of influence Earman and Glymour's arguments have had, and how damaging their claims have been to scientific integrity. Stanley summarizes this rather nicely on his page 328: 'It is not unusual today to meet physicists who have accepted Collins and Pinch's (really Earman and Glymour's) critique of the 1919 results. They will talk about error bars and bias. This version of the story has become a kind of scientific folklore, passed around at water coolers. *The Golem* was a fantastically successful scholarly publication … and its arguments have filtered out into many other books in many different disciplines. Very few physicists repeating this version of the story know where it came from, though. They would probably be quite shocked to learn that they were passing along arguments aimed at undermining the very foundations of their field.'

Prior to 1980, descriptions of the 1919 eclipse results largely repeated the formal report conclusion. While there were questions raised about the precision of the results (see appendix B), there was no suggestion that the data selection had been affected by bias. That claim was introduced by Earman and Glymour for the first time. Commentators discussing the issue of bias and/or the robustness of the results refer either directly to their article or to the popularization of its findings in *The golem*, where the authors repeat the arguments with exaggeration and present the findings, unchecked, as newly established

---

evidence. *The golem* certainly succeeds in making Earman and Glymour's assessment of the expedition results a new point of reference.[29]

Doubts about the integrity and scientific objectivity of the 1919 eclipse team in general, and Eddington in particular, whether popular belief or 'water cooler tales', have become ubiquitous. To the best of our knowledge, they all derive from the Earman and Glymour analysis discussed here. No other study has analysed the original 1919 eclipse data and claimed on that basis that the original data selection involved bias. Their arguments remain influential, remain cited and remain used at face value in academic publications to illustrate or strengthen a philosophical point with reference to 1919.

As Kennefick, Stanley and others point out, more recently scientists themselves have done the protagonists of the 1919 expeditions no favours, and suggestions of (confirmation) bias are repeated not just in casual conversation but also in publications referring to the 1919 results.

The most obvious example is a reference in Stephen Hawking's popular 1988 book, which we discuss further in appendix B. Another is a comment by Francis Everitt in a rather more obscure publication dating back to 1980, to which a referee generously drew our attention.[30] Everitt's comments rely on a very strange (mis)interpretation of the results, which does not make any (scientific) sense against the background of the 1919 expedition report. This seems to be an example, indeed, of a scientist getting the numbers wrong and deriving inappropriately critical conclusions from his interpretation. It is worth noting, however, that in spite of Everitt's strong scientific credentials, his clearly incorrect description of the 1919 data analysis was not picked up in the scientific community.

Everitt's and Hawking's critical comments are testimonies of an increasing general disenchantment in the scientific community with the accuracy of the measurements and the reliability of the data the 1919 (and other) expeditions had generated. We believe that these concerns match an evolution from 'Einstein or Newton' to precision tests of General Relativity, and can in part be connected to a 1960 publication by Harald von Klüber; we discuss this further in appendix B. Von Klüber's comprehensive summary of all—successful and unsuccessful—eclipse expeditions and his critical presentation of their methodology provides good grounds for the belief that eclipse expeditions have not generally provided very reliable data to make any strong statements about Einsteins's

---

29  Collins and Pinch, *op. cit.* (note 7), afterword, pp. 157–158, and at p. 171.

30  A referee generously draws attention to a comment by C. Francis Everitt from 1980 'Others [i.e. other plates] again from an astrographic camera at Sobral gave a very-reliable looking measurement of $0.93 \pm 0.05$ arcsec—the scaling coefficient must have been wrong, so they were thrown out though the evidence for them was much better than that for the $1.61 \pm 0.30$ arcsec measurement at Principe. It is impossible to avoid the impression—indeed Eddington virtually says so—that the experimenters approached their work with a determination to prove Einstein right. Only Eddington's disarming way of spinning a yarn could have convinced anyone that here was a good check of General Relativity.' C. W. F. Everitt, 'Experimental tests of General Relativity: past, present and future', in *Physics and contemporary needs* (ed. Riazuddin), pp. 529–555 (Plenum Press, New York, 1980). The article by Everitt is a reprint of the introductory part of an earlier article: C. W. F. Everitt, J. A. Lipa and G. J. Siddall, *Precision Eng.* **1**, 5–11 (1979). See also footnote 77 of Stanley (2003), *op. cit.* (note 13), and footnote 181 of Stanley (2007), *op. cit.* (note 13). Everitt's statement is quite remarkable. He invents a very small uncertainty for the Sobral astrograph result, with no explanation. He mis-characterizes the detailed discussion of systematic errors in the Dyson, Eddington and Davidson report by using the terms 'very-reliable looking' and 'evidence for them was much better' for the Sobral astrograph plates. Everitt's claims were not noted at the time—the first citation of this paper seems to be in 2003, and, more recently, several references that tend to treat the statements as robust: Schindler, *op. cit.* (note 14), p. 97 has 'Earman and Glymour's assessment receives independent support from the physicist C. W. F. Everitt', and see note 345 in Stephen G. Brush, *Making 20th century science* (Oxford University Press, New York, 2015). The 'Eddington as storyteller' theme has a substantial literature of its own. See, for example, Alistair Sponsel, 'Constructing a "revolution in science": the campaign to promote a favourable reception for the 1919 solar eclipse experiments', *Br. J. Hist. Sci.* **35**, 439–467 (2002). Note also our discussion of Almassi in note 13. A point for future analysis is the way the community has seemingly uncritically accepted Everitt's comments at face value in spite of their very evident unexplained inconsistencies with the supporting evidence he cites.

Theory of Relativity. Resulting misgivings, such as Hawking's, about the success of the 1919 expeditions, take no account of the exceptional nature of the 1919 eclipse or the actual purpose of the experiment. They do provide an echo to Earman and Glymour's claims, but are not themselves attempts to evidence bias and lack of scientific integrity.

## Discussion

Analysis of experimental data in 1919 by experienced astronomers was published and accepted as technically valid by the astronomical community of the time, after they had considered and discounted many possible extraneous sources of systematic error. The consequences for theoretical physics were profound, so the results were carefully scrutinized. We have investigated how scientific expertise acknowledged by expert peer-review in 1919 came to be invalidated and transformed into a story of bias many decades later. We note that accusations of scientific bias, exaggerated in later literature, lead back to the re-analysis of the scientific data published 60 years later by philosophers of science Earman and Glymour, who claimed that the verification of General Relativity did not rest on the evidence presented but was influenced by bias.

We have repeated both the original and the 1980s analyses to understand how an accepted scientific result, one of the higher public profile experiments of the twentieth century, could be so easily questioned and undermined. We have re-analysed the published 1919 eclipse data to check the analysis of the time. The original 1919 analysis is statistically robust, with conclusions validly derived, supporting Einstein's prediction. The rejected third data set is indeed of such low weight that its suppression or inclusion has no effect on the final result for the light deflection, though the very large and poorly quantified systematic errors justify its rejection. We have re-analysed the Earman and Glymour 1980 paper, scrutinized their statistical analysis and conclude that it is not methodologically sound.

Earman and Glymour's publication falls into the philosophical context of analyses of models of the scientific method. In conclusion to their re-analysis of the 1919 expedition results, they claim to have found evidence against 'science as a model of objectivity and rationality'.

The main purpose of their re-analysis is to strengthen a philosophical point, not a scientific one. Their 'discovery' makes no difference to the credibility of Einstein's Theory of General Relativity. In the authors' own words, that 'theory, so far as we know, still holds the truth about space, time and gravity'; and yet the arguments they use to discredit the scientific method then applied have had influence and consequence that go beyond the philosophical debate from which they emerged—especially once they reached the wider public domain unchecked.

While acknowledging the raison d'être of the philosophical investigation that forms the background to their arguments, we show that Earman and Glymour failed to distinguish between and appreciate different areas of expertise, and have conflated arguments that belong to different fields of investigation.

Examining the particular case of the 1919 British eclipse expeditions as an example in their philosophical enquiry, they sought to underpin their main argument with 'new' scientific evidence. Their underlying mathematical analysis provided Earman and Glymour with a very powerful premise to an argument that became influential in both academic and popular publications, producing much stronger evidence for bias than supposition and speculation ever could. In doing so they assumed themselves to be experts both in the field of philosophy of science and in the field of scientific data analysis.

The epistemological and methodological weakness we have identified in their argument was overlooked or ignored by their readers and readily exploited by later authors (most notably in *The golem*). Unquestioned as new experts, by at least a significant sub-set of the community, their claim that the results of the 1919 expedition were affected by bias was allowed to establish itself as evidence-based.

As Kennefick and Stanley's discussions both show, and as one referee generously points out, it is fair to clarify that 'not everyone has, as it were, been "taken in"' by Earman and Glymour's (and Collins and Pinch's) arguments.[31] However, the background of Earman and Glymour's 'findings' adds difficulty to arguments that the decisions made by the scientists in 1919 were *not* affected by bias—and even those who remain unconvinced by the authors' claims cannot simply ignore them. Unexamined and unrefuted, they remain available as 'scientific' evidence to those who wish to strengthen their own philosophical arguments about bias with reference to the 1919 expeditions. Without Earman and Glymour's re-analysis of the results, or indeed with that re-analysis refuted, the 1919 decision-making process followed by Dyson and Eddington is much more clearly definable. It is an open question how much notice would have been taken of Earman and Glymour's claims at all without the re-analysis of the results underpinning them.

With the powerful premise to their argument now invalidated, the question must be asked: what really is left of Earman and Glymour's philosophical claim to have found in the 1919 expeditions evidence against science as a model of objectivity and rationality? Against the revised background we have provided, the motives for the decisions made by the scientists in 1919 move back more firmly into the realm of objectivity and rationality, and a philosophical argument for bias in 1919 would have to be constructed differently. Arguably, perhaps, the motives for the 1980 re-analysis and for the popularization of its findings may then also warrant further discussion.

Objectivity and rationality clearly are important, not only in science but in any scholarly endeavour that broadly aims to be also a quest for the truth. Proper critical review by experts in all aspects of the results presented in any given research or experiment remains the most powerful tool towards achieving this as a standard in academic research. It is also a responsibility on the part of the peer community, given the importance, the influence and the power of references to academic expertise and expert arguments in the wider community of non-experts.

Earman and Glymour's influence on Eddington's reputation of bias is just one example where more objectivity and rationality could have led to a quite different story being told.

ACKNOWLEDGEMENTS

The least squares fitting of the published astrometry presented in appendix A was implemented as part of a Cambridge undergraduate project by Isaac Webber, completed in May 2019. The fitting used the MATLAB lsqr routine to solve the linear equations and the results were checked by the authors. We thank Peter Coles for drawing our attention to the articles by Bauer, mentioned in note 17, and a referee for noting the article by Everitt. Helpful comments by the referees have substantially improved the presentation of this article.

---

31  See note 30.

This work received partial support from the European Union H2020 programme through grant number 730890, OPTICON (Optical Infrared Coordination Network for Astronomy).

### APPENDIX A: 1919 ECLIPSE STATISTICAL RE-ANALYSIS

#### Available data

All data used in this analysis are those published in the original eclipse report.

Measuring precise stellar positions at the time involved differential small-angle astrometry. An image of a star field is compared with a different image of the same field, overlaid in a measuring machine. The difference in centres of the nearly overlapping images are then measured, quantifying any stellar movement. For convenience, a reversed ('scale') plate of the same field can be used as an intermediate step. What is required are plates of the star field both with (day-time eclipse plates) and without (night-time comparison plates) the eclipsed Sun. To check for systematic errors due to different optical setups, 'check plates', images of another (non-eclipse) star field taken with the same telescope, were obtained at both Oxford and Principe.

The coordinates of the stars were measured relative to a scale plate and compared with later night-time comparison plates of the eclipse field; the scale plate is only intermediary. Key positional changes are:

1. Differences in scale value ('magnification'), orientation and centring between the plates.
2. Distortion of the star field by gravitational deflection.
3. Any other distorting cause.
4. Accidental errors of observation.

Secondary influences are differential atmospheric refraction and aberration, which are corrected using calculated values.

Dyson commented on the data measurement at the Royal Astronomical Society's Discussion on Relativity meeting in December 1919:[32] 'I wish to remove misconceptions which some Fellows may have with reference to the observations. In linear measurement the quantities dealt with are of the order of $1/100^{th}$ of a millimetre. These displacements are small, but Astronomers familiar with stellar photographs will, I am sure, support the statement that quantities of this order are readily measured, if only good photographs are secured.' That is, he was explaining to the non-expert that the techniques involved here were standard. The image quality and image shift are illustrated in figure 1.

#### Sobral 4-inch

Seven stars were bright enough to be measurable on the seven cloud-free (out of eight total) eclipse photographs obtained with this telescope. The plates were measured independently by Charles Davidson and Mr Furner at the Royal Greenwich Observatory by placing them emulsion-to-emulsion against a fifteenth reference plate of the star field, observed through the glass plate.

---

32  Sir Frank Dyson, 'Discussion on the Theory of Relativity', *Mon. Not. R. Astr. Soc.* **80**, 106 (1919).

The intermediate ('scale') plate is used purely as a differential convenience and does not contribute to the final results. Astrometric measurement records the difference in position of each star from its image on the reference plate. This process minimizes measurement error, and, as described in the eclipse report, there was no significant contribution to the error budget from the measurement process. All plates were measured independently in the (orthogonal) directions right ascension and declination. All these differential positions were published in the report, corrected for (atmospheric) differential refraction and (astronomical) aberration, and are available for re-analysis. These same locally-appropriate corrections were made to all the data from the three telescopes.

*Sobral astrographic*

Eighteen exposures were obtained with the Sobral astrographic telescope during the eclipse, of which 16 were of measurable depth. In July eight comparison plates were obtained. The eclipse plates were measured directly against the comparison plates without the need for a separate reference plate. The eclipse plate images were of poor quality, so special-purpose analysis was used. The data published in the report are 'only measured in Declination, as the Right Ascensions were of little weight'.

Summary results, but not the full set of position measures, derived from the analysis are published for all 16 plates (two measured against two separate comparison plates, so 18 values in total), with the number of stars measurable varying from 7 to 12 (eclipse report, table IX). These summary results list, for each plate, the derived zero-point orientation offset ('d'), the apparent difference of scale between the eclipse and comparison plates ('e') and the derived constant of the gravitational displacement, both as measured ('α') and the value of gravitational displacement at the Sun's limb in arcseconds. These results cannot be re-derived without the original relative position data. However, measured relative displacements in declination are published for five stars on the 16 plates, which can be re-analysed.

*Principe astrographic*

Only two eclipse plates were deemed satisfactory, with five stars measurable on each. The remaining 14 were affected by cloud.[33] Night-time check plates (i.e., of a different star field) were obtained, with the comparison and matching check plates obtained from Oxford with the same telescope prior to the eclipse.

The data published are the differential positions of the five stars between each eclipse plate and two comparison plates, providing four comparisons with five stars in each. Differential measures of Principe versus Oxford check plates are also published to determine the measurement accuracy, and to determine any change of the plate scale between Oxford and Principe exposures.

---

33  A third plate, Plate U, was reduced at Principe to give a deflection of 2.9 arcsec with a probable error of 0.9 arcsec. This was later rejected as too uncertain to be used. It is this plate that is the second mentioned in Eddington's letter to his mother from Principe indicating preliminary agreement with Einstein's prediction.

*Analysis method*

Techniques for analysing (differential) astrometric images were already developed at the time of the eclipse, and involved solving (first order) equations for each star of the form

$$Dx = ax + by + c \tag{1}$$
$$Dy = dx + ey + f, \tag{2}$$

with $x$ and $y$ the positions of each star (relative to an arbitrary set of axes), $Dx$ and $Dy$ the position differences of the stars between the two plates being measured and a, b, c, d, e, f being a set of (plate) constants. This allows for a scaling, rotation and enlargement between images. These equations are modified in the present application to include terms to account for an additional change in position due to any gravitational deflection, to become

$$Dx = ax + by + c + \alpha Ex \tag{3}$$
$$Dy = dx + ey + f + \alpha Ey. \tag{4}$$

Here $Ex$ and $Ey$ are the anticipated gravitational displacements, scaled so that α represents the deflection at the Sun's limb in arcseconds.

For the Sobral 4-inch, data solution of the 14 equations (one for each star in each direction) is straightforward. The eclipse report used the method of 'Normal Equations'; we use least squares. The other two data sets are more complex. The Principe astrograph data were reduced (by Eddington) using an iterative process, and using the scale plate as an intermediate step, where again we use least squares. The Sobral astrograph data proved problematic. The eclipse images are out of focus and distorted, while night-time plates, with no adjustment made, were in focus. At the time, the observers noted this as being probably due to uneven heating of the (coelostat) mirror, and doubted that robust results could be derived.[34] The issue, apparent to the experienced observers, relates to a possible scale change of the telescope in addition to the mirror distortion affecting image quality. The eclipse report provides summary results from a fit of equation (4) above to the full set of declination measures, solving for all of d, e and α. In addition, fits both by Normal Equations and by least squares to the data for the five brightest stars are reported.

The results from the report and our analysis of the derived gravitational deflection, corrected to the limb of the Sun, are in agreement, with the small differences consistent with rounding errors (table A1). The calculations at the time were evidently mathematically robust.

*Accuracy, uncertainty, precision, dispersion*

It is impossible to deduce the significance of a measurement without knowing an appropriate measure of its uncertainty. Much of the discussion in the report is dedicated to determining the several contributions to the final uncertainty associated with the mean gravitational deflections determined above. The most important for this discussion are the random errors, dominated by the accuracy of individual star positions, the reliability with which independent measures can be averaged together to improve mean accuracy and the systematic errors, which were dominated by determination of scale.

---

34  Dyson *et al.*, *op. cit.* (note 2), footnote on p. 309.

Table A1. Values of the mean gravitational deflection from the eclipse report and from modern re-reduction.

| Telescope | Reduction method | Report value (arcsec) | Recalculation (arcsec) |
|---|---|---|---|
| Sobral 4-inch | Normal equations—declination | 1.94 | |
| | Normal equations—right ascension | 2.06 | |
| | Normal equations—combined | 1.98 | |
| | Least squares—combined | | 1.99 |
| Sobral astrograph | Normal equations all data | 0.86 | |
| | Normal equations five stars | 0.93 | 0.93 |
| | Least squares five stars | 0.99 | 0.95 |
| Principe astrograph | Normal equations iterative | 1.61 | |
| | Least squares | | 1.61 |

It is worth emphasizing that all position measurements were made by experienced astronomers using state-of-the-art measuring machines. The limiting factor is the quality of the photographic plates. We also note that the system used in the report for expressing errors was the standard at the time, the probable error, which functions like a modern standard deviation. Increasing a probable error by a factor of 1.48 gives the equivalent standard deviation for error values that follow the Gaussian (normal) distribution.

We recall the standard definitions, appropriate to distributions that follow the normal distribution.

The standard deviation (dispersion) of a set of numbers is:

$$\sigma = \sqrt{\left(\frac{1}{N-1}\sum (x_i - \bar{x})^2\right)}.$$

The uncertainty of the mean when combining $N$ values is:

$$\sigma_\mu = \frac{\sigma}{\sqrt{N}}.$$

The weighted mean, and its uncertainty, when combining values each with their own error estimate is:

$$\mu = \frac{\sum (x_i/\sigma_i^2)}{\sum (1/\sigma_i^2)}; \quad \sigma_\mu^2 = \frac{1}{\sum (1/\sigma_i^2)}.$$

The total uncertainty on a mean value is the sum in quadrature of the random error and the systematic error. Because random and systematic errors behave differently, they are usually kept separate, though in the report they are combined, when available, into a single final value.

*Sobral 4-inch*

The derived 'probable error' of 0.12 arcsec (corresponding to a standard deviation of 0.18 arcsec) is calculated in the report from 'the accordance of the separate determinations'. This value is consistent with the listed differences presented as 'observed minus calculated' between the deflection predicted by General Relativity and the observations, in the traditional astronomy style, on page 309. These have standard deviations of 0.11 and 0.12 arcsec. These are consistent with our least squares fit, which provides a dispersion of 0.325 arcsec for the seven plates, corresponding to an uncertainty on the mean of 0.13 arcsec. The scale-comparison analysis in the report shows systematic errors are unlikely to be larger than the random errors, so that quadrature addition confirms the report uncertainty on the mean of 0.18 arcsec as a reasonable value for later use.

*Principe astrographic*

The report analysis gives a final probable error of 0.30 arcsec (deliberately a round number) with a significant systematic contribution. This corresponds to a standard deviation of 0.45 arcsec. The dispersion of the four results in the report is 0.22 arcsec. Our least squares fit corresponds to a dispersion of results of 0.46 arcsec. Given the correlated errors and systematics described in the report, this cannot simply be scaled to an uncertainty on the mean reduced by a factor $\sqrt{(N-1)} = 1.7$, to a value of 0.3 arcsec. The report value of 0.45 arcsec is a reasonable conservative value.

*Sobral astrographic*

No uncertainties are quoted in the report for the Sobral astrograph data. As is emphasized, this is because of the likelihood of large systematic errors, and the 'considerable range in the deduced values'. Additionally, as is apparent from report page 311, there was a systematic change in the observing system during the eclipse. Plates 15 to 18 are statistically significantly discordant with plates 1 to 12 (the difference in deflection between the plate sub-sets is 100%). The set of measures does not follow a normal distribution. The internal dispersion of the measures of all available stars on the 18 data sets (two plates measured twice) is 0.47 arcsec with the mean deflection of 0.86 arcsec. The restricted five-star data set gave a mean of 0.93 arcsec, and with a different assumption on scale a limb deflection of 1.52 arcsec. Our least squares fit may be compared with the mean value of 0.99 arcsec with no quoted error in the report. We derive a mean of 0.95 arcsec with a dispersion about that mean of 0.90 arcsec. Given the non-Gaussian error distribution and unknown systematics noted above, there is no model-independent way to use this dispersion to derive a standard error on the mean. This large dispersion does confirm the report conclusion to neglect this result.

We have now for the Sobral 4-inch and Principe astrograph telescope data sets estimated uncertainties on the quoted mean deflection, and a measure of the internal dispersion of all three data sets. These are summarized in table A2. Note that the Sobral astrograph errors do not include possible large systematic errors.

It is now straightforward to calculate, in modern terminology, the statistical significance of the 1919 eclipse results (table A3). It is also possible to combine the several telescope results to quantify the impact of including/rejecting the Sobral astrograph results. It is important to remember that the error distributions are affected to some degree by systematics, so these significance levels are indicative only. So, for example, the results from the Principe astrograph are consistent with the Newtonian prediction with 10% probability, the Sobral 4-inch results are consistent only with the Einstein deflection, while the Sobral astrograph results do not allow any distinction between any of the three hypotheses. We discuss these significance levels in the main text.

Table A2. Errors and dispersions for the three data sets.

| Telescope | Probable error (report) (arcsec) | Standard deviation (equivalent) (arcsec) | Least squares dispersion (arcsec) | Mean (arcsec) | Error on mean (arcsec) | Internal dispersion (arcsec) | Relative weight |
|---|---|---|---|---|---|---|---|
| Sobral 4-inch | 0.12 | 0.18 | 0.33 | 1.98 | 0.18 | 0.11 | 25 |
| Principe astrograph | 0.30 | 0.45 | 0.46 | 1.61 | 0.45 | 0.22 | 4 |
| Sobral astrograph | None | None | 0.90 | 0.95 | 0.9 | 0.47 | 1 |

Table A3. Comparison of the agreements of each telescope with the three theoretical hypotheses. The $P(z)$ values are the probabilities of observing a result at least as extreme as the prediction due to the Einsteinian, Newtonian or null (no deflection) model (in either tail) of a normal distribution with mean and standard deviation of the appropriate telescope.

| Telescope analysis | Deflection (arcsec) | Einstein | | Newton | | Null | |
|---|---|---|---|---|---|---|---|
| | | z | P(z) | z | P(z) | z | P(z) |
| Sobral 4-inch | 1.98 ± 0.18 | 1.28 | 0.201 | 6.17 | 0.000 | 11.0 | 0.000 |
| Principe astrograph | 1.61 ± 0.45 | −0.31 | 0.756 | 1.64 | 0.100 | 3.58 | 0.000 |
| Sobral astrograph | 0.95 ± 0.9 | −0.89 | 0.373 | 0.09 | 0.929 | 1.06 | 0.290 |

Table A4. Indicative outcomes from forming weighted combinations of the three separate telescope results. The offsets are in units of standard deviations. A result at the 3–$\sigma$ level is highly significant. Note that these results are indicative, not robust.

| Combination | Weighted mean (arcsec) | Offset from Einstein | Offset from Newton |
|---|---|---|---|
| Sobral 4-inch + Principe astrograph | 1.93 ± 0.17 | 1.06$\sigma$ | 6.2$\sigma$ |
| Sobral 4-inch + Sobral astrograph + Principe astrograph | 1.90 ± 0.16 | 0.94$\sigma$ | 6.4$\sigma$ |
| Sobral 4-inch + Sobral astrograph with error = 0.48 arcsec + Principe astrograph | 1.82 ± 0.16 | 0.44$\sigma$ | 5.9$\sigma$ |

One may also form suitably weighted combinations of the results from the different telescopes (table A4). Note, however, that this assumes independent normal-distributed results, without significant systematic error sources, so that these values are indicative but certainly not statistically robust.

A result at the 3–$\sigma$ level is nominally significant. The six-sigma results here indicate very high degrees of significance. Thus, all possible combinations strongly prefer the Einstein deflection over the Newtonian value. The third row adopts (artificially) the internal dispersion for the Sobral astrograph data set as if it were a standard deviation including all error sources; that is, accepting the Earman and Glymour numbers at face value, and so testing their conclusion that this would change the results of the whole experiment. This gives that data set significantly higher weight. Interestingly, the agreement with the Einstein prediction is strongest in this case.

APPENDIX B: CONTEMPORARY AND LATER CRITICISM

Contemporary queries were raised to consider other possible physical causes that might generate a deflection. These were considered, and their refutations listed, by Dyson in 1921 and others. A similar list of possible alternative physical effects was considered, and rejected, in the analysis of the next eclipse, that of September 1922.[35]

That is evidenced both in the special edition of *Nature* in 1921 devoted to relativity, which has an article by Dyson on the eclipse results from May 1919, and in von Klüber's later (1960) review, which has a comprehensive list of relevant references.[36]

H. N. Russell, a very eminent and experienced astrometrist and astronomer, re-reduced the 4-inch plates' measures and provided an independent confirmation of the report analysis and its accuracy, specifically the quoted uncertainties for the 4-inch results.

Some later astronomers did derive different results from the published eclipse data, but without general acceptance. For example, Danjon re-analysed all of 1919, 1922-I, 1922-II and 1929 eclipse data, and in every case derived a limb deflection of 2.06 arcsec, with surprisingly small scatter. He did this by introducing a free-parameter related to plate scale to force the observations to match the $1/r$ law exactly; this paper did not make an impact on the field.[37]

W. W. Campbell, the leader of the Lick Observatory 1914 and 1922 eclipse expeditions, is often quoted as being critical of the data selection in 1919, from a 1923 lecture in which he gives a 'history of our previous efforts', commenting that the 'logic of the [data selection] situation does not seem entirely clear' and concluding 'a point of interest is that the British observers were the first to say … that confirmation should be sought at the eclipse of Sept 21 1922'.[38] We suggest the reader considers the comments in their original context and decides for themselves if this is criticism of the data selection or, rather, emphasis that two

---

35 See note 11. Also see notes 17 and 20.

36 *Nature* **106**, pp. 781–813 (1921); Dyson, *op. cit.* (note 19). See also note 17.

37 A. Danjon, 'Le déplacement apparent des étoiles autour du soleil éclipsé', *J. Phys. Radium* **3**(7), 281–301 (1932). (doi:10.1051/jphysrad:0193200307028100).

38 W. W. Campbell, 'The total eclipse of the Sun, September 21, 1922', *Publ. Astr. Soc. Pacific* **35**, 11–44 (1923). The relevant section, p. 19, reads: 'The two cameras used in Brazil, where fortunately the weather conditions were excellent, gave Einstein coefficients differing in amount. With the astrographic camera, a duplicate of that employed in Africa, were secured 16 plates, recording 6–11 stars each. The images were not in good focus, apparently because the heat of the Sun's rays falling upon the plane coelostat mirror, which reflected the stellar rays into the horizontal camera, caused distortion. However, all of these plates were measured. The results for the coefficient varied from 0″.00 to 1″.28. Their mean value, 0″.86, agreed well with the half Einstein effect, 0″.87, predicted in 1911. The second camera gave photographs in excellent focus, and their measurement and discussion yielded an Einstein coefficient of 1″.98, in good accord with the full, or double, effect 1″.75, resulting from the generalized theory of relativity. The observers attributed by far the greater weight to the last result; in fact, assigning very little weight to the first of the Brazilian results. Professor Eddington was inclined to assign considerable weight to the African determination, but, as the few images on his small number of astrographic plates were not so good as those on the astrographic plates secured in Brazil, and the results from the latter were given almost negligible weight, the logic of the situation does not seem entirely clear. A point of interest is that the British of the three 1919 data sets lacked precision, thus by implication strengthening the need for Campbell's planned confirmatory second experiment. Campbell's early history of expeditions was corrected by Perrine.[39]

It has also been argued that the 1919 eclipse results, while accepted at the time, were criticized by astronomers soon after. Hentschel, for example, concludes: 'The experiments previously held to be decisive for the … general theory, such as light deflection … proved on the contrary to be completely inadequate, because their accuracy had been overrated, while the systematic errors were too large for the instruments available at that time to surmount. This is true in retrospect also of those experiments that had been accepted by the majority of Einstein's competent contemporaries as key verifications of relativity theory, such as Eddington's and Crommelin's measurements of light deflection in the sun's gravitational field of 1919.'[40] Hentschel's conclusion is heavily based on later 'reanalyses of the 1919 and 1922 results by Freundlich *et al.* (1931)'. Hentschel is an outlier in accepting Freundlich's re-analyses. They were not considered credible at the time, or later, and were extensively criticized, as Hentschel himself describes. As we show in this paper, the accuracy of the 1919 experiment had not been overrated. The systematic errors for the 1919 and 1922, but crucially not the 1929, eclipses were under control.

A further and complementary key factor to take into account is the later change in the community approach, from verification of an effect to precision test—where the exactitude of the value measured for the deflection is the focus of the enquiry rather than the verification of a theory consistent with the mathematical derivation of 1.75 arcsec. We have already highlighted the difference between the two approaches in our discussion of Earman and Glymour's criticism of the trichotomy. Keeping the two questions distinct is essential. The focus on precision measurement inevitably highlights the limits of early work, and we believe that later (1980s) comparison and conflation of the two approaches may have been a factor in the tendency to doubt the 1919 results. Precision measurement was at any rate the explicit motivation for the Potsdam 1929 expedition led by Freundlich.

It is worth emphasizing that no-one expert in precision astronomical measurement raised any concerns about the 1919 (or 1922) measurements, or the data selection. Those consistently opposed after about 1930 were not expert astrometrists. It may seem surprising, but this also applies to Freundlich.[41] Freundlich's experience was wide-field measurements, in which absolute measures are needed and control of systematics is key;

---

observers were the first to say, in view of the fundamental importance of the general subject, that confirmation should be sought at the eclipse of September 21, 1922.'

39  C. D. Perrine, 'Contribution to the history of attempts to test the theory of relativity by means of astronomical observations', *Astron. Nachricht.* **219**, 281–282 (1923)—the 1912 first dedicated light-bending eclipse test has recently been described in detail by Luis C. B. Crispino and Santiago Paolantonio, 'The first attempts to measure light deflection by the Sun', *Nat. Astron.* **4**, 6–9 (2020) (doi:10.1038/s41550-019-0995-5), and in S. James Gates and Cathie Pelletier, *Proving Einstein right: the daring expeditions that changed how we look at the universe* (Public Affairs, New York, 2019).

40  Klaus Hentschel, 'Erwin Finlay Freundlich and testing Einstein's Theory of Relativity', *Arch. Hist. Exact Sci.* **47**, 143–201 (1994), at p. 192; Klaus Hentschel, *The Einstein tower: an intertexture of dynamic construction, Relativity Theory and astronomy* (Stanford University Press, Stanford, 1997). See also Lewis Pyenson, 'Einstein's early scientific collaboration', *Hist. Stud. Phys. Sci.* **7**, 83–123 (1976). The 1929 eclipse results are published in E. F. Freundlich, H. von Klüber and A. Von Brunn, 'Ergebnisse der Potsdamer Expedition zur Beobachtung der Sonnenfinsternis von 1929, Mai 9, in Takengon (Nordsumatra). 5. Mitteilung, *Über die Ablenkung des Lichtes im Schwerefeld der Sonne*' *Z. Astrophys.* **3**, 171–198 (1931). See also Crelinsten, *op. cit.* (note 13), and Crelinsten, *op. cit.* (note 11) at note 11, ch. 11.

41  Crelinsten, *op. cit.* (note 13) at p. 16, and Crelinsten, *op. cit.* (note 11) at footnote 40.

hence his series of works with von Klüber and von Brunn, his design and construction of a sophisticated self-calibrated eclipse camera system (used in 1929) and his emphases on absolute scale calibration of the measurements, which von Klüber continued in his 1960 review.

A detailed explanation of the need for consistency of telescope alignment, including deliberately retaining misalignments, specifically for the two astrographs used in the 1919 eclipse, is provided by the astrometrist Derek H. P. Jones.[42] This study shows explicitly that the Freundlich approach to calibration did not influence the expert community of astrometrists, at the time or later—and remained distinct from it. The difference between Freundlich's approach and that of expert astrometrists is also apparent in the correspondence between Freundlich on the one hand and Ludendorff and Trümpler on the other.[43]

Von Klüber's 1960 review of all earlier light-deflection eclipse experiments contains a summary table, which became widely copied. The review itself is uncritical of published results, and does not consider statistical significance, while supporting the Freundlich *et al.* approach to large scale calibration. A reader without experience in observational astronomy could easily get the impression not only of discordant results, but also that the measuring tools and techniques were *per se* inadequate at any given time and irrespective of the particular experiment conducted, and would have led to measurements too imprecise to provide robust scientific evidence. This sort of criticism found its way into more contemporary literature, including Stephen Hawking's 1988 bestseller: 'It is ironic, therefore, that later examination of the photographs taken on that expedition showed the errors were as great as the effect they were trying to measure. Their measurement had been sheer luck, or a case of knowing the result they wanted to get, not an uncommon occurrence in science.'[44]

It is not impossible that Hawking's comments are derived from a textbook by his PhD supervisor, Denis Sciama, who uncritically reproduces the eclipse list from von Klüber (1960) and makes very generalizing comments on eclipse measurements and their precision with regard to testing deflection.[45]

---

42  D. H. P. Jones, 'The Greenwich and Oxford astrographic telescopes 1958–1987', in *Mapping the sky. Past heritage and future direction: IAU Symposium 133* (Kluwer, Dordrecht, 1988), pp. 33–38.

43  Freundlich *et al.*, *op. cit.* (note 40); H. Ludendorff, 'Über die Ablenkung des Lichtes im Schwerefelde der Sonne', *Astron. Nachricht.* **244**(16), 321 (1932), note his table 2; J. Jackson, 'The deflection of light in the Sun's gravitational field', *Observatory* **54**, 292–296 (1931); Robert Trümpler, 'Die Ablenkung des Lichtes im Schwerefeld der Sonne. Mit 3 Abbildungen', *Z. Astrophys.* **4**, 208 (1932), with an English summary in *Science* 20 May 1932, and reprinted in *Publ. Astr. Soc. Pacific* **44**, 167 (1932), with a footnote correcting comments on the 1922 Lick results by W. W. Campbell. Freundlich *et al.* also criticized the Lick 1922 eclipse results, claiming (ironically given the circumstances) systematic position-dependent residuals. W. W. Campbell, in a footnote on p. 172 of Trümpler 1932 (this note) replied: 'The extensive newspaper publicity given in several countries to the Potsdam criticism of the results obtained by the Lick Observatory at the 1922 Australian eclipse has been in such form as to cause the readers to conclude that the Lick observers paid no attention to the run in the residuals of the check star field. This is not only unfortunate, from our point of view, but is also diametrically the opposite of the facts in the case.'

44  Stephen W. Hawking, *A brief history of time* (Bantam, Toronto, 1988), at p. 32. Hawking's error was corrected by P. A. Wayman and C. A. Murray, 'Relativistic light deflections', *Observatory* **109**, 189–191 (1989), though a publication less well-known than his book.

45  Denis Sciama, *The physical foundations of General Relativity* (Doubleday, New York, 1969) at pp. 70–71. Sciama states 'Ironically enough, we shall see that Einstein's prediction has not been verified as decisively as once believed', then comments 'there are several cases in astronomy where knowing the "right" answer has led to observed results later shown to be beyond the power of the apparatus to detect'.

We note that while contemporary criticism was refuted or not taken seriously by the community of experts, later criticism of the 1919 results based on their lack of precision may have originated from sources that disregarded a number of factors crucial to understanding the 1919 expeditions' success, such as the exceptional nature of the eclipse, the parameters of the experiment and the actual expertise available at the time.

In summary, sceptical responses to the 1919 eclipse results are of four quite different types: (1) non-gravitational explanations for the *undisputed* apparent deflection measured (contemporary); (2) criticism by less expert astrometrists, most prominently Freundlich (1930s); (3) post-1980 generalized scepticism of eclipse precision, occasionally incorrectly focused on a specific eclipse, which we tie at least in part to a summary article by Freundlich's colleague von Klüber; (4) post-1980 claims of bias originating from within a philosophical enquiry into the scientific method and using the 1919 expeditions as a historical example, but not based on scientific evidence.